\documentclass[pre,twocolumn,superscriptaddress,showpacs,amsmath,amssymb,reprint,groupedaddress]{revtex4-1}

\usepackage{amsmath, amssymb}

\usepackage{graphicx} 
\usepackage{times,mathptmx}
\usepackage[usenames]{color} 
\usepackage{soul}

\begin{document}

\newcommand{\comA}[1]{\textcolor{black}{#1}}
\newcommand{\comAnke}[1]{\textcolor{red}{#1}}
\newcommand{\comN}[1]{\textcolor{magenta}{#1}}

\newcommand\s{\dot{\gamma}}
\newcommand\sm{\dot{\gamma}_{mean}}
\newcommand\slo{\dot{\gamma}_{local}}
\newcommand\sa{|\dot{\gamma}|}

\newcommand\al{\textit{et~al.}\ }
\newcommand\ec{\textit{E. coli }}

\title{Lagrangian 3D tracking of fluorescent microscopic objects \comA{in motion}}

\author{T. Darnige}
\author{N. Figueroa-Morales}
\author{P. Bohec}
\author{A. Lindner}
\author{E. Cl\'{e}ment}

\affiliation{Laboratoire Physique et M\'{e}canique des Milieux H\'{e}t\'{e}rog\`{e}nes, CNRS- ESPCI Paris, PSL Research University, Universit\'{e}s Pierre et
Marie Curie and Denis Diderot, Sorbonne-Universit\'{e}s, 10, rue Vauquelin, Paris, France.}


\begin{abstract}

We describe the development of a tracking device, mounted on an epi-fluorescent inverted microscope, suited to obtain time resolved 3D Lagrangian tracks of fluorescent \comA{passive or active micro-objects} in micro-fluidic devices. The system is based on real-time image processing, determining the displacement of a x,y mechanical stage to keep the chosen object at a fixed position in the observation frame. The z displacement is based on the refocusing of the fluorescent object determining the displacement of a piezo mover keeping the moving object in focus.  Track coordinates of the object with respect to the micro-fluidic device, as well as images of the object are obtained at a frequency of several tenths of Hertz. This device is particularly well adapted to obtain trajectories of motile micro-organisms in micro-fluidic devices \comA{with or without flow}.

\end{abstract}

\maketitle


\section{Introduction}

In their seminal work Berg and Brown have shown in 1972 that tracking individual micro-organisms in 3D was a determinant step to understand bacterial motility and chemotactic response to chemical gradients \cite{berg1972chemotaxis}. At that time the device built by Berg, designed to perform the tracking and 3D trajectory reconstruction of bacteria, was conceptually and technically outstanding \cite{berg1978tracking}.

Since then, other techniques of 3D tracking have emerged.
For example, the development of fast piezo scanning has led to high speed image stack acquisition techniques. These techniques were used to reconstruct, by post processing, the swimming trajectories of micro-organisms \cite{corkidi2008tracking}. This method, though very powerful to simultaneously obtain the trajectories of multiple objects, finds its limitation if the object speed is high. Also, it is strongly dependent on computer memory capacities which limits the possibility of tracking over long times.
Alternative techniques using defocussing rings of fluorescent particles have been used to access the $(x,y,z)$ position of micro-objects and a large number of tracks at the same time  \cite{wu2006collective}. The limitations here are the complexity of the post-processing, increasing with particle concentration, and the limited range of $z$ exploration. Past the year 2000, there was a surge in the development of  holographic 3D particle reconstruction techniques (see for example a review by  Memmolo et al. \cite{memmolo2015recent} and refs inside). Using post-processing, these tracking methods lead to the 3D reconstruction of a few simple trajectories of objects as Brownian spherical colloids \cite{cheong2010strategies} or bacteria \cite{saglimbeni2014digital}. This method has recently led to important advances in the understanding of sperm swimming and chemotaxis \cite{su2012high,jikeli2015sperm}.

However, all of these recent 3D tracking methods are of Eulerian type, meaning that the volume of observation is fixed in the laboratory reference frame. It is important to notice that such Eulerian methods may inherently lead to a systematic statistical bias. Particles with very persistent trajectories are likely to leave the volume of observation faster than particles undergoing for example a diffusive motion.

\begin{figure}[!htb]

	\includegraphics[width=\linewidth]{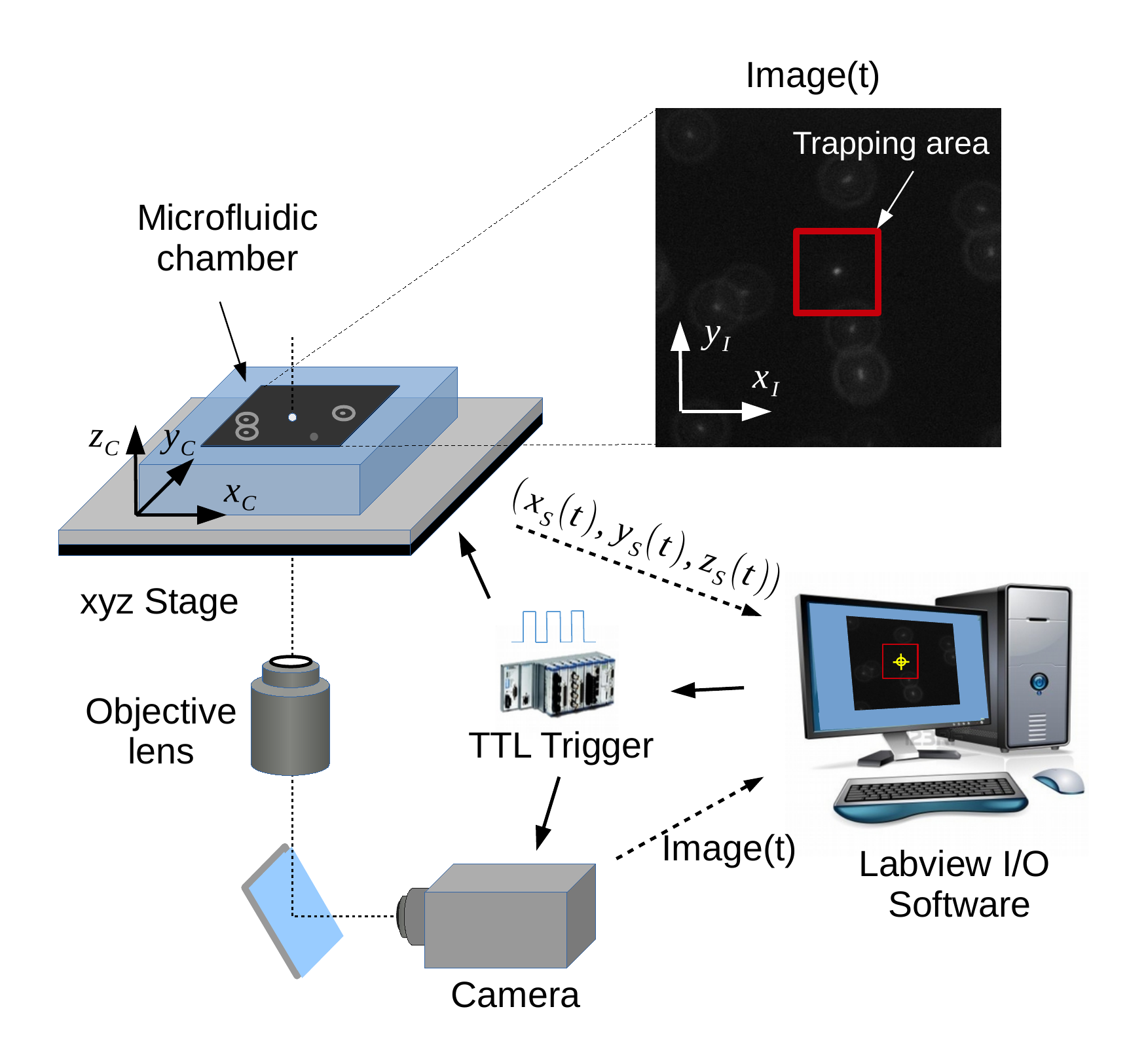}

 \caption{Experimental set-up. The tracking system consists of two superimposed stages mounted on an inverted microscope. The horizontal $x,y$ position is controlled mechanically and the $z$-position via a piezo-electric mover. The targeted fluorescent particle is visualized in a ``trapping area'' using a CCD camera. \comA{The stages and the camera are triggered by a National Instrument TTL trigger module to synchronize the image acquisition and the stage displacement.} The image is transferred to a Labview program processing the information. The program records the current $x,y$ and $z$ positions and commands the mechanical and the piezo stages to move to a new position such as to keep the particle close to the ``trapping area'' and in focus. The outputs are (i) a record of the $x,y,z$ positions and (ii) a video of the particle in its environment during the Lagrangian tracking.}
 \label{fig:setup}
\end{figure}

\begin{figure}[!htb]
	\includegraphics[width=1\linewidth]{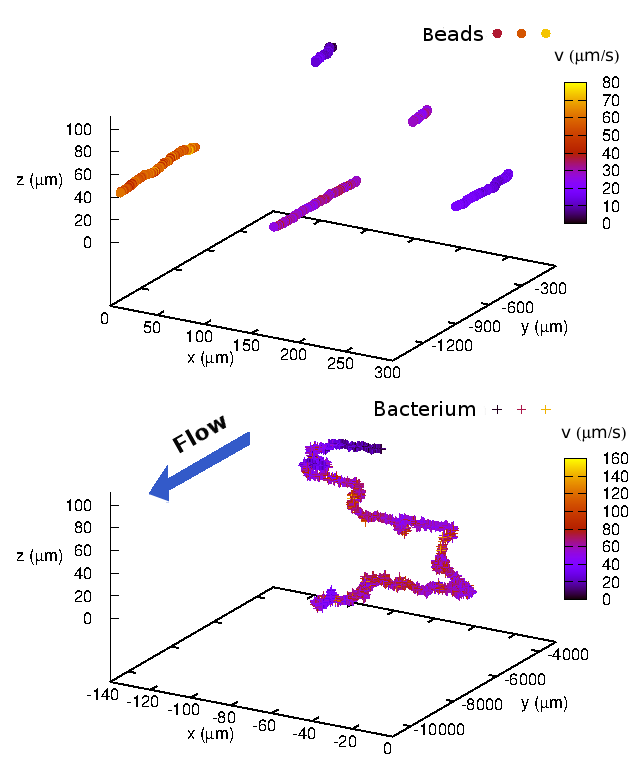}

 \caption{3D tracks of latex beads (top) and a bacterium (bottom) under flow for the same experiment in a microfluidic rectangular channel. The colour code stands for local velocities in micrometers per second. The micro-channel spans from $z=0\mu m$ to $110\mu m$ in height, and the width from $x = -300\mu m$ to $x = 300\mu m$. The length is approximately 150mm \comA{and the flow was imposed by applying a pressure difference. The particles are latex beads of diameter $1\mu m$. The bacterium here is a wild-type strain of {\it E. coli} (RP 437), transfected with a Yellow fluorescent plasmid.
 The bacteria are grown overnight in the rich culture medium M9G plus antibiotics (Chloramphenicol) at \comA{$30 ^oC$} and then harvested (optical density $= 0.5$ at $595 nm$). The cells are then washed and suspended in a minimal medium favorable for their motiliy and seeded with the latex beads. }}
 \label{fig:track3D}
\end{figure}

Lately, increasing interest into the swimming dynamics of micro-organisms has made the development of reliable Lagrangian tracking techniques using modern tools particularly important again  \cite{liu2014helical,turner2016visualizing}. In this paper we describe the technical details of a Lagrangian 3D tracking method that extends the seminal work of Berg \cite{berg1978tracking} using modern visualization tools and data treatment possibilities. Our technique relies on a time resolved refocusing of a fluorescent \comA{spheroidal micrometric} object (Fig. \ref{fig:setup}) and has the advantage of providing not only the full 3D trajectory of the particle but also a direct image of the tracked particle in its environment. \comA{It can be used to track fluorescently labelled swimming micro-organisms as \textit{E. coli} bacteria or fluorescent passive objects like silica or latex particles in quiescent fluids as well as in imposed flows (Fig. \ref{fig:track3D}).} From the 3D tracks of bacteria in quiescent fluids one can for example obtain insights in the run and tumble dynamics of different bacteria strains. 3D tracks of bacteria under flow can reveal the particular interaction of the latter with a given flow geometry.  Following passive tracer particles under flow can be used, for example, for an accurate determination of the velocity profile in micro-fluidic devices.
In addition, a unique feature of the technique described here, is the possibility to track individual bright spots in a crowded environment of opaque objects. It can for example be used to follow a specific fluorescent bacterium in a suspension of none fluorescent bacteria, in order to characterize the features of collective motion.

\section{Experimental set-up}

The set-up is sketched in Fig. \ref{fig:setup}. It is composed of an inverted microscope (Zeiss-Observer, Z1) with a high magnification objective ($100 \times/0.9$ DIC Zeiss EC Epiplan-Neofluar), a $xy$ mechanically controllable stage with a $z$ piezo-mover from Applied Scientific Instrumentation (ms-2000-flat-top-xyz) and a digital camera ANDOR iXon 897 EMCCD. \comA{The $x,y$ axis range of travel is 120mm$\times$75mm, the $x,y$ resolution is 22nm and the $x,y$ maximum velocity is 7mm/sec. The $z$ axis range of travel is $500 \mu m$ and the $z$ resolution is 8nm}.  \comA{The stage and camera are both triggered by a National Instrument TTL trigger module (NI 9402 Bidirectionnal digital input/ouput, 4 channels, LVTTL) to get synchronized images and stage positions.}
The camera output and the stage input/output signals are connected to a multi-threaded Labview software, capable of reassigning the stage position by a real-time feedback loop, keeping one particle in focus in the central visualization region (see Fig. \ref{fig:setup}). The ANDOR camera is very sensitive (\comA{with a single photon detection capability through enhanced cooling and a} pixel size of $16$ microns) and designed to visualize even weak fluorescent signals. It is thus relatively slow, working nominally at 30 fps on a 512 $\times$ 512 $pix^2$ matrix. Here, we present the performances achieved at a faster tracking speed of $80 Hz$ reducing the spatial resolution to 128 $\times$ 128 $pix^2$. The tracking limitations come essentially from the $z$ exploration range, limited essentially by the working distance of $150 \mu m$ of the $100 \times$ objective.

From the coordinates of the stage in the laboratory reference frame and the coordinates of the position of the fluorescent object in the corresponding image, we obtain the three-dimensional trajectory. A video of the object and its surrounding is also recorded, providing a direct visualization of its projection in the $x,y$ plane.

In Fig. \ref{fig:track3D}, we present some tracking results obtained with this technique, for passive fluorescent particles and for a fluorescent \textit{E. coli} bacterium in a rectangular microfluidic cell under flow. \comA{For the passive tracer particles of diameter $1\mu m$ straight trajectories of constant velocity are observed. The tracer velocity is a function of the height in the channel due to the Poiseuille flow profile typical for microfluidic devices. The microorganism has an ellipsoidal body of semiaxes of typical dimensions $1\mu m$ and $2\mu m$. One can see that its trajectory is composed of series of almost straight paths separated by short abrupt random changes of direction. This dynamics is known as \textit{run and tumble} \cite{berg1972chemotaxis}. The typical run duration is close to $1s$ and during such a run the bacterium is observed to swims at $10$ to $30 \mu m/s$. A tumble takes typically about $0.1$s during which the swimmer velocity is strongly reduced \cite{berg1972chemotaxis}.}

Our technique can also be adapted to an objective of lesser magnification or using a much faster (though less sensitive) SCMOS technology. The rate of transfer of visual information, its processing and the reaction of the mechanical and piezo-components of the $x,y$ and $z$ stages have in all cases to be adapted to the possibilities of the material components used. Nevertheless, the method described here remains valid despite changes of visualization techniques.

\subsection{General algorithm}

In the following, we describe the general detection algorithm.
First, when a fluorescent particle reaches the trapping area, the user can decide by clicking on the mouse to start the tracking process. The first action of the detector is to determine the $(x_I, y_I)$ position of the targeted object on the image and to put it in the center of the image by moving the $x,y$ mechanical stage. Then the system proceeds by scanning very rapidly in $z$ to get a defocussing refocusing video that will immediately be analyzed. This first step allows to determine several initialization parameters that will be used subsequently to specify the detection functions. Then, the algorithm starts to work in the nominal mode. At every iteration step of the detection algorithm, an image $Image(t)$ (i.e. a 128x128 matrix of 14 bytes per pixel), as well as the current coordinates of the stage \comA{$(x_S(t),y_S(t), z_S(t))$} are obtained and transferred to the Labview program. Then the target position for the object in the image reference frame is determined and a 3D move is performed by the stage in order to keep the object in the focus volume. Then, at $t + dt$, the next image exposure takes place. For each iteration, the object position and the image are recorded. In the following, we describe the principles of $x,y$ and $z$ detection and thereafter the $x,y$ and $z$ stage motions. Finally, we address the question of the tracking performances.

\subsubsection{x,y detection}

\begin{figure}
\begin{tabular}{l l}
\textbf{a} \hspace{4cm} & \textbf{b}\hspace{4cm} \\
\end{tabular}

	\includegraphics[width=\linewidth]{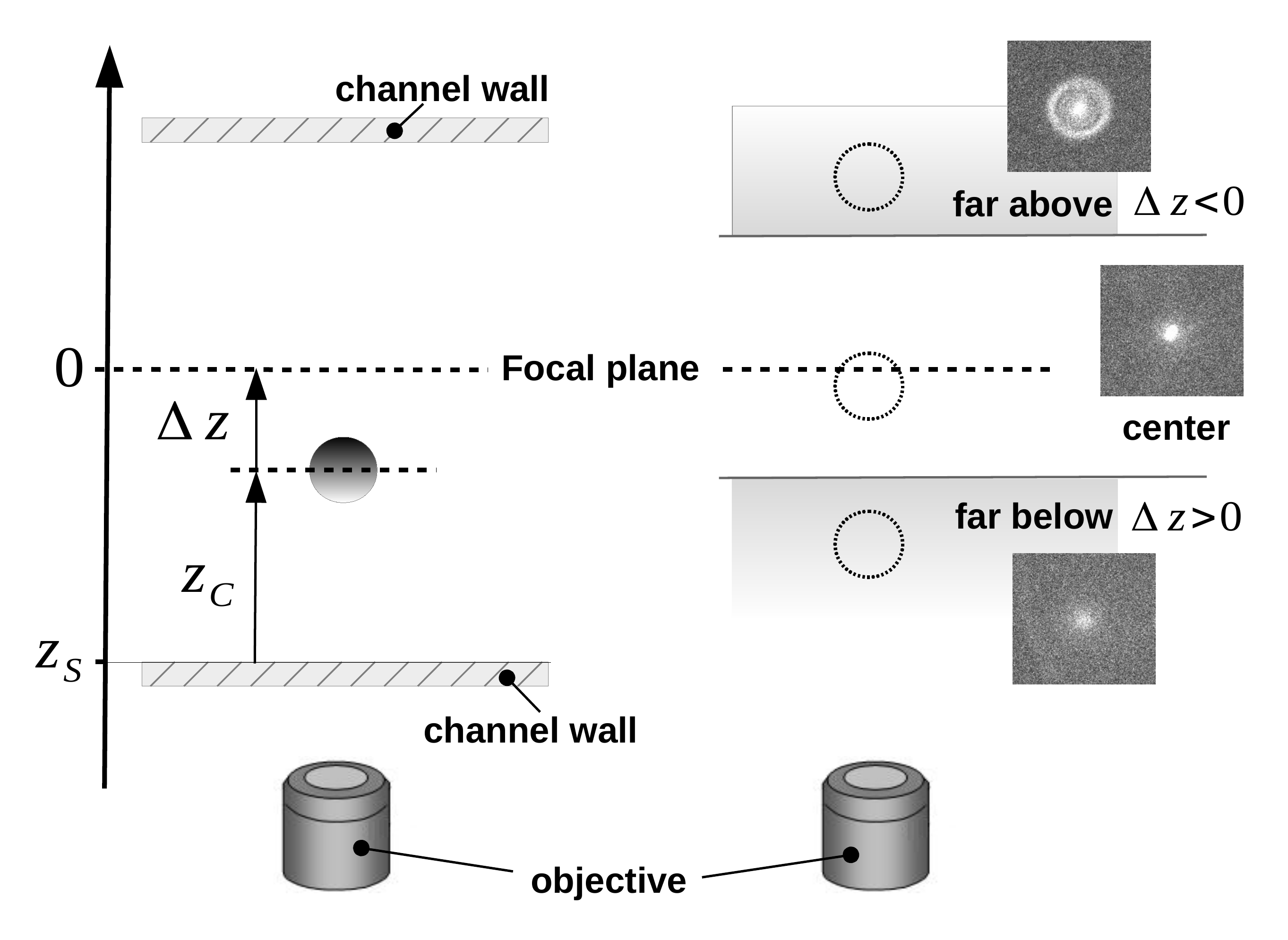}
 \caption{ \comA{(a) Sketch illustrating the positions of the particle, stage and focal plane in the reference frames associated to the laboratory and to the microfluidic channel. The correction $\Delta z$ is measured from the object to the focal plane position. The origin of the laboratory reference frame coincides with the focal plane. (\textbf{b}) Illustration of the different patterns obtained for different positions of the object with respect to the focal plane.}}
 \label{fig:sketch_Z}
\end{figure}

\begin{figure*}[!htb]
\begin{tabular}{l l l}
\textbf{a} \hspace{5.5cm} & \textbf{b}\hspace{5.5cm} & \textbf{c} \hspace{4cm} \\
 \\
\end{tabular}
\centering	\includegraphics[width=\linewidth]{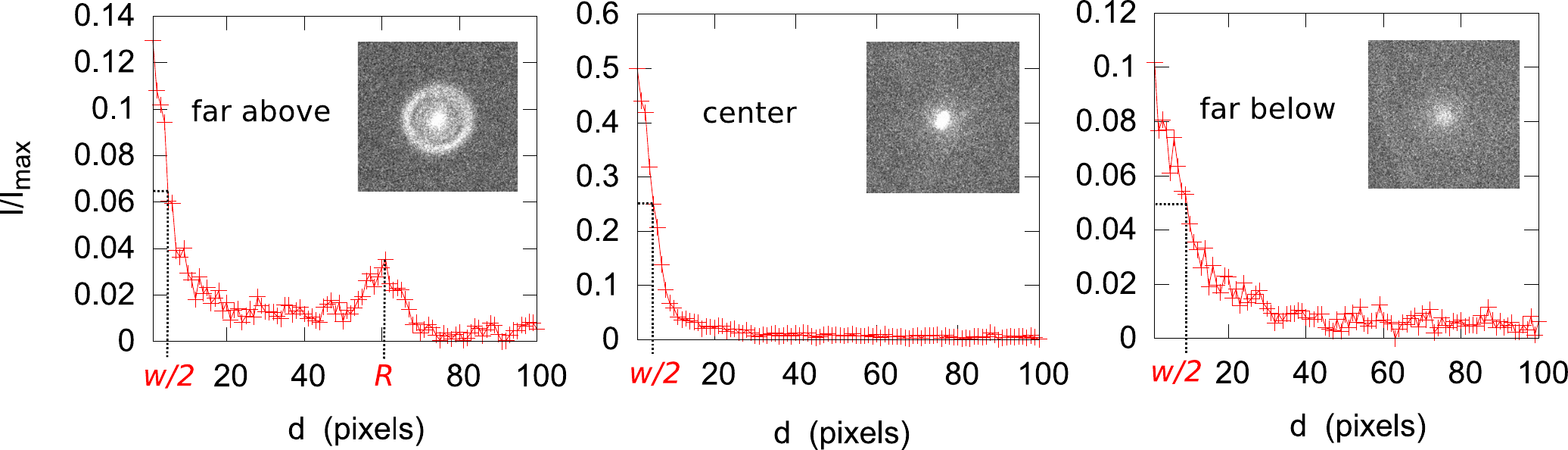}
 \caption{Normalized intensity profile as a function of the distance $d$ to the center of the bacterium. $I_{max}$ is the absolute intensity maximum when the particle is in focus. \comA{The intensity profile is qualitatively different when the object if far above the focal plane (panel \textbf{a}), close to it (panel \textbf{b}) or far below the focal plane (panel \textbf{c}). The half-width $w/2$ of the particle (size at half intensity) is indicated in each case in red letters, as well as the ring radius $R$ in the region far above.}}
 \label{fig:IvsR}
\end{figure*}

To track the targeted object in two-dimensions, we use a standard image processing method. The object coordinates in the reference frame of a micro-fluidic chamber: \comA{$(x_C, y_C)$}, are composed of the stage coordinates in the laboratory reference frame when the image was exposed \comA{$(x_S, y_S)$}, and the  coordinates of the object within the image \comA{$(x_I, y_I)$. Note that the image and laboratory reference frames are equivalent, except for an offset that we can we can set to zero at the beginning of the experiment. The relevant coordinates of the object, from a physical point of view, are those in the channel reference frame as they reflect the motion of the object with respect to its environment, i.e., the microfluidic device.} They can be obtained as:
\comA{
\begin{eqnarray}
 x_C(t) = x_I(t) - x_S(t) \\
 y_C(t) = y_I(t) - y_S(t) \mbox{.}
 \label{x_chamber}
\end{eqnarray}
}

The object does not need to be in the center of the image during the detection process. Using the positions $(x_I(t), y_I(t))$ and $(x_S(t+dt), y_S(t+dt))$, we define a search area in $Image(t+dt)$ (extracted Region Of Interest, ROI) around the position $(x_I(t), y_I(t))$ where the object is expected to be in the next image. This ROI is pre-processed using a gaussian filter and a local threshold, to get a binary image with a background correction such as to obtain a set of binarized objects. The objects are filtered to keep only those in a predefined size range. The positions of their geometrical centers are determined. We keep as $(x_C(t+dt), y_C(t+dt))$ the closest object to the former position $(x_C(t), y_C(t))$. One can easily introduce some memory (as we will explain later for $z$) in order to obtain continuity of trajectories for example after a collision between two similar particles. However, the present method turns out to be quite efficient and permits a computational time of under $5ms$ and thus a margin of $5ms$ for the $z$ determination and the stage motion. This is compatible -in principle- with a tracking at a rate of 100 frames per second. We thus decide not to use a memory function for the $x,y$ detection as it would decrease our time performance.

\subsubsection{z detection}

The principle for the $z$ position detection differs significantly from the detection of the horizontal $(x,y)$ coordinates.  It is based on an optimized search for a vertical position suited to keep the moving object in focus. \comA{Note that on our set-up the objective is fixed and as the focal plane is located at a constant distance from the objective, the location of the focal plane is fixed in the laboratory reference frame. During the tracking the stage is moving in the $z$ direction to keep the object in focus. When the particle moves up, the stage moves down, and vice-versa, to keep the distance between particle and objective constant (Fig. \ref{fig:sketch_Z}(\textbf{a})).}

\comA{The vertical coordinate $z_C$ of the object in the channel reference frame can be determined from the coordinate of the stage $z_S$ in the laboratory reference frame and a correction $\Delta z$ accounting for the distance between the object and the focal plane. Fig. \ref{fig:sketch_Z} (\textbf{a}) represents these variables. We chose the origin of the laboratory reference frame such as to coincide with the focal plane, leading to $z_C = -z_S - \Delta z$, where $\Delta z$ is measured from the particle to the focal plane and can be positive or negative and is zero when the object is in focus. A positive displacement of the particle has to be compensated with a negative stage displacement of the same magnitude and vice-versa ($d z_C = -d z_S $), in order to keep the particle in focus, as explained above.}

\comA{In the following we describe the detailed procedure for the $z$ detection for the example of an \textit{E. coli} bacterium. The method and calibration remains identical for other types of fluorescent particles of dimensions close to $2 \mu m$ or below that are not too elongated (typically an aspect ratio below 3.) }

\comA{The object width is minimal when it is in focus and grows when it gets further away from the focal plane. Our algorithm keeps the object in focus by minimizing the apparent object width $w$. This allows to account for the photo-bleaching of the object producing a temporal decrease of the apparent size of the object. In addition it turns out to be a very robust criterion, nearly insensitive to the noise induced by the presence of other bright objects that may temporally appear in the background.}

\begin{figure}[!htb]
	\includegraphics[width=\linewidth]{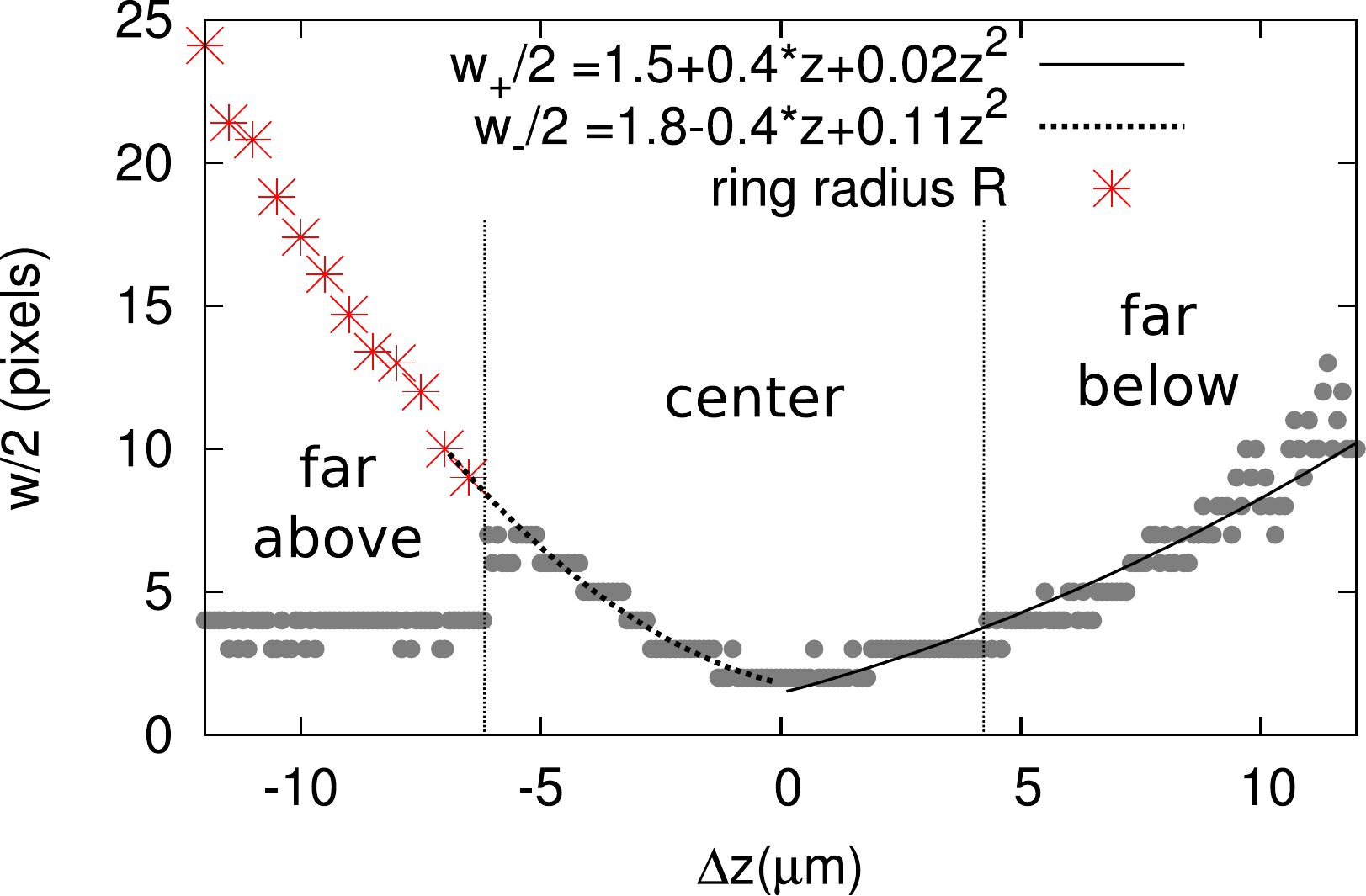}	
 \caption{Half width $w/2$ \comA{and ring radius $R$ of a bacterium} as a function of the distance between the object and the focal plane. The particle is in focus at $\Delta z=0$.}
 \label{fig:width_Z}
\end{figure}

\comA{For the $z$ detection, we define three different regions regarding the position of the object compared to the focal plane   (see Fig. \ref{fig:sketch_Z} (\textbf{b})): (i) the region ``far above'', where the particle is above the focal plane ($\Delta z < 0 $), (ii) the region ``far below'', where the particle is far below the focal plane ($\Delta z > 0 $) and (iii) the ``center'' region, where the particle is close to the focal plane and $\Delta z $ is small and can be either positive or negative. }

\comA{The characteristics of the obtained images differ from one region to another.}
On Fig.  \ref{fig:IvsR} we show the mean radial intensity profiles \comA{and the half-width ($w/2$) of a typically fluorescent \textit{E. coli} bacterium} obtained from images in these three different regions. The average intensity ($I(d)$) profile, averaged over 360 degrees, is computed around the position of the particle $(x_I(t), y_I(t))$ as a function of the distance to the center ($d$). As a convention, we compute the width of the object ($w$) as twice the distance from the center to the point where the intensity has dropped to one-half of the central intensity maximum. \comA{Although bacteria are not perfectly spherical, for their not very elongated bodies an average radial intensity profile turns out to work very well. } We now describe how we determine $\Delta z$ from the images in the different regions.

\emph{\comA{Far above~-~}} This region is characterized by the presence of rings around the particle \comA{as can be seen from Fig. \ref{fig:IvsR} (\textbf{a}), where the ring radius $R$ is indicated. To obtain a calibration for $R$ as a function of the distance to the focal plane, a z-scan of a bacterium attached to the bottom of the chamber is performed. Typically, the rings appear when the object is further than $6\mu m$ above the focal plane. The ring size $R$ as a function of the distance to the focal plane $\Delta z$ is approximately linear up to a distance of $20\mu m$ (Fig. \ref{fig:width_Z}). A calibration of the form $\Delta z = c R$ can be performed by adjusting the data. With this calibration, the measured value of $R$ can be used to estimate $\Delta z$ and thus the distance from the focal plane. }


\emph{\comA{Far below~-~}} In this region the particle appears as a fuzzy white dot of low intensity with no ring. If the maximum object intensity $I(t)$ is lower than a given fraction of the currently stored $I_{max}$ value (this coefficient is typically set to $0.4$ but can be adjusted), the algorithm recognizes that the \comA{object is far below the focal plane}. Again a z-scan is performed to measure the width $w$ as a function of $\Delta z$. We then establish a calibration function $w_{+}(\Delta z)$ for the width of the particle as a function of its distance to the focal plane.
The correction will be calculated as
$\Delta z(w, w_{min}) = (w - w_{min}) \left. \frac{d \Delta z}{d w_{+}}\right|_{w}$.
Here $w_{+}(\Delta z)$ is a second order polynomial represented in Fig. \ref{fig:width_Z} by the solid line on the right side of the curve. Typically this region starts at a distance between object and focal plan of around $4\mu m$.

\emph{Center region ~-~} In the center region the z scan of the bacterium fixed at the bottom plane reveals two curves for the object's width as a function of $\Delta z$: one curve for $\Delta z<0$ ($w_{-}(|\Delta z|)$) and one for $\Delta z>0$ ($w_{+}(|\Delta z|)$, above the object) where the center $\Delta z = 0$ corresponds to the local minimum. The curves are fitted in Fig. \ref{fig:width_Z} and they are almost mirror-symmetric close to the center. In the center region it is not possible to know from the image whether the focal plane is slightly above or slightly below the object.
We can therefore only work with one detection function and we have chosen the right branch $w_{+}(|\Delta z|)$, working reasonably well in both cases. We relate the object's position to the measured width $w$ by means of the inverse function. The absolute value will be $|\Delta z|(w, w_{min}) = (w - w_{min}) \left. \frac{d z}{d w} \right|_{w}$. The sign of $\Delta z$ can not be know from a single image and only the comparison of the value of $w$ between two consecutive images sets the sign of $\Delta z$.

\comA{Note that the function $|\Delta z|(w, w_{min}) = (w - w_{min}) \left. \frac{d z}{d w} \right|_{w}$ used for the center region and the region far below is identical. It is however useful to define the region far below (using as a criterion $I(t)<0.4 I_{max}$ as explained above) as in the far below region the sign of $\Delta z$ is know from a single image.}

\begin{figure}
	\includegraphics[width=1\linewidth]{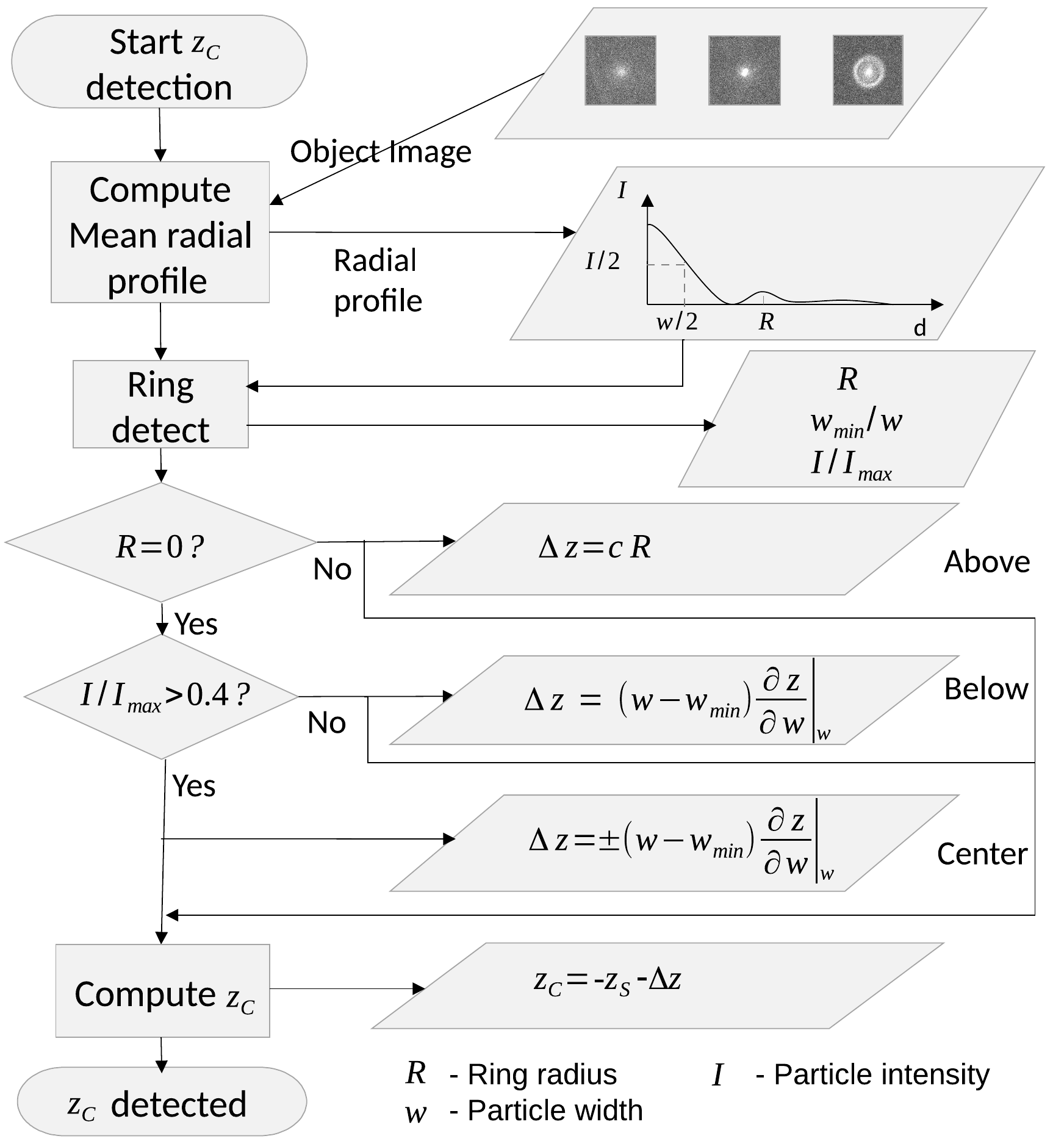}
 \caption{Flow diagram of the $z$ detection. }
 \label{fig:z_detection_flow_diagram}
\end{figure}

\subsubsection{$x,y$ and $z$ stage displacement}

During the tracking initiation procedure, the fast $z$-scan yields a sequence of images typically separated by $0.5\mu m$ around the object. For every image, the object width $w$  is determined and thereafter its minimum value $w_{min}$. From the image where $w_{min}$ is minimal we also measure and store the maximum intensity $I_{max}$ and the vertical position $z_{C}$. At the end of the sweep, the stage goes to the best $z$ coordinate and the tracking variables are fully initialized.

Note that all along the tracking process, the detected minimum width $w_{min}$ and the central intensity $I_{max}$ at the instant of the minimum width detection, are kept in memory. Whenever a  smaller width is encountered through the tracking process, this value becomes the new $w_{min}$ and the new maximum intensity replaces the current $I_{max}$ value. The regular update of these parameters allows us to keep track of photo-bleaching which produces, on the long run, a drastic intensity decrease and a reduction in the apparent object size.

For the $x,y$ stage motion, a proportional integrative derivative (PID) algorithm is used and empirically tuned, to perform a progressive motion of the detection zone. This avoids to directly couple detection and displacement. Such a direct coupling would keep the object perfectly centered in the image but could potentially introduce mechanical perturbations in particular for abrupt movements of the object. For our $x,y$ detection the object does not need to be in the center of the image and we thus prefer to use a progressive displacement of the stage.

For the $z$ stage displacement, when the detected region is far below or far above, the stage is moved with a value $\Delta z_S(w, w_{min}) = \Delta z(w, w_{min})$ such as to reach the center region in one step.

In the center region the sign of $\Delta z$ is not know from a single image. Only the comparison between two successive images permits to determine wether the particle is above or below the focal plane. If the last step was successful (thus corresponding to a decrease in $w$), we keep this direction of stage displacement and if not we invert it. As a result, the position of the object fluctuates around the focal plane. The stage is moved at a displacement amplitude $\Delta z_S(w, w_{min}, \vec{v}_z)$ where we take into account a second order contribution to the position, incorporating in the computation the displacement of the object between two consecutive images. In this way, $\Delta z_{S}(w, w_{min}, v_z) = \Delta z(w, w_{min}) - k(w, w_{min}) v_z \Delta t$, where $v_z$ is the object velocity along $z$, computed in real time from the comparison of the two pervious images. The coefficient $k(w, w_{min})$ is positive and between 0 and 1. The value 0 is used for $w$ close to $w_{min}$ and 1 is used when $w$ significantly differs from $w_{min}$. The coefficient $k(w, w_{min})$ is obtained from an empirical extrapolation function $k(w, w_{min}) = \frac{1}{2}\left[\tanh(\frac{0.95-w_{min}/w}{0.05})+1\right]$\comA{. This function permits to progressively react to a decrease of the focus quality. When the focus is good it is not necessary to consider the velocity of the object $v_z$ for the determination of $\Delta z_S$. However, when the focus become less good and the object leaves the focal plane fast, the term involving its velocity becomes important as a bigger additional correction to the $z$ stage displacement is needed to keep the particle in the center region.}

Note that a descending bacterium, leaving the focal plane towards the bottom of the channel, is more easily followed than a bacterium swimming to the top of the channel. This is due to the fact that the far below region is reached at a distance of about $4\mu m$ whereas the far above region is only reached at a around $6\mu m$ from the focal plane. As soon as the particle has reached the far below region it can be moved back into the focal plane in one step.

As said previously, the sign of $\Delta z(w)$ depends on the change of $w$  (decrease or increase) when compared to successive images. Due to the buffer of the camera, images are obtained with a delay of two periods of sampling. As a consequence the most recently captured image is not a direct consequence of the last displacement of the stage.
\comA{To keep the correct synchronization between stage displacements and image analysis, we have reduced the effective sampling and displacement rates in $z$ to 1/3 compared with $x$ and $y$. In this way, the $z$ coordinate is updated at a frequency of $26.7Hz$.}

\subsubsection{Backlash correction}

On our set-up the $x,y$ motion of the stage is performed using two lead screws, one for each axis. An inherent feature of such a mechanical device is the  existence of a backlash, meaning that the screws need some finite back-rotation to restart a reverse motion.
For our system the backlash is of the order of $1 \mu m$. Since our stage controller uses rotary encoders, this repetitive error is not visible from the recorded stage data. So in the  post-processing, we use a correction algorithm in order to remove this error from the tracking data. Replacing the rotary encoder by a linear optical encoder, returning the real position of the stage, would make this correction unnecessary. Note that no backslash occurs for the $z$ displacement which is piezoelectric and thus provides a high precision positioning at the nanometer scale.

 Now we describe how we correct our data for the mechanical backlash error in $x$ and $y$. In general, high precision stages perform very precise displacements by following an algorithm for motion which is designed to avoid backlash problems. Before each move, the stage goes to an intermediate position far away and thus always approaches the targeted position from the same direction. This is certainly not suitable for particle tracking where the stage has to follow a moving particle in real time. Here we compensate for the stage backlash in a post-processing treatment of the data. Note that this step is crucial and special care has to be taken when building a 3D tracking device in order to avoid spurious data recording due to the backlash problem.

\begin{figure}[!htb]

    \includegraphics[width=1\linewidth]{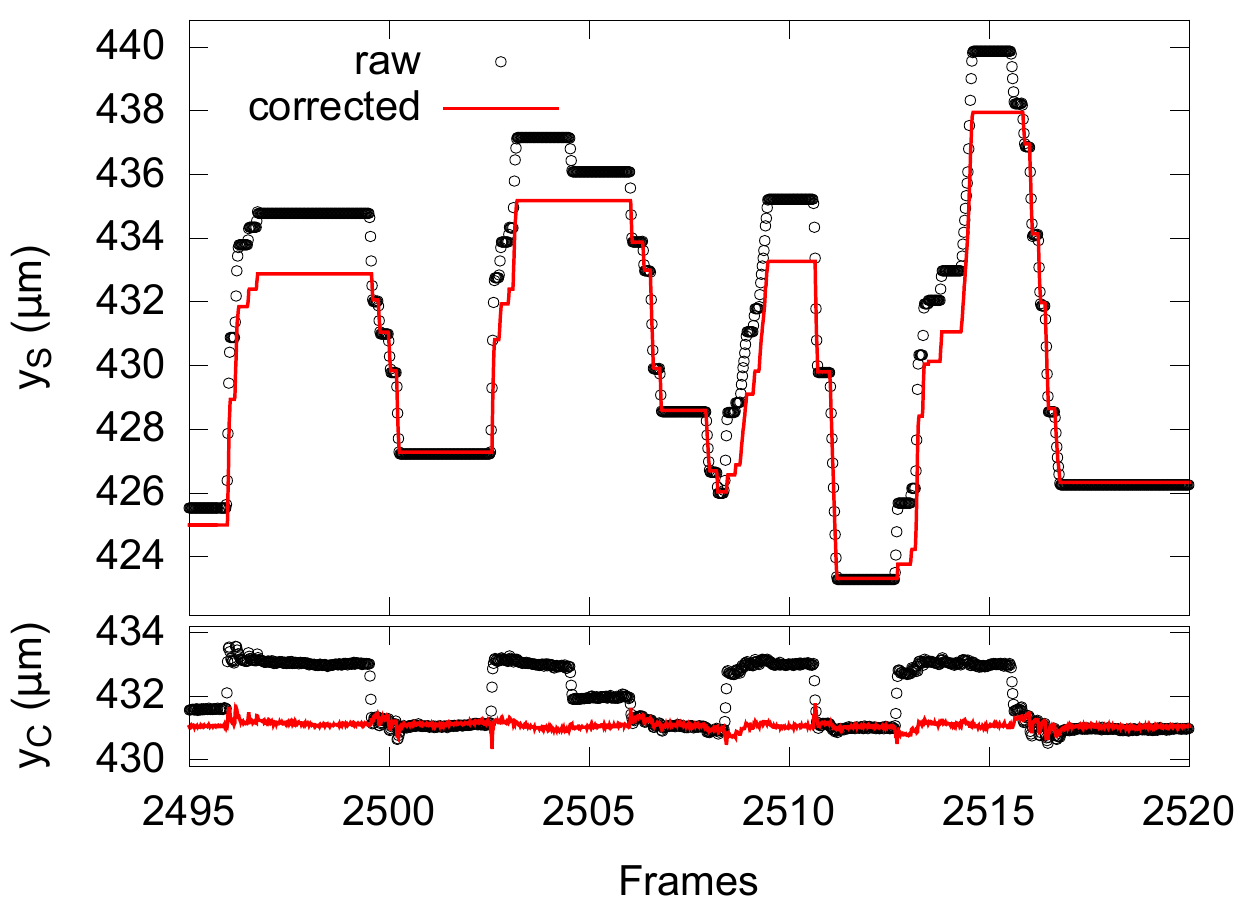}

 \caption{$y$ coordinates of the stage (top panel) and particle position (bottom panel). Black circles are the raw measurements and the red line represents the same data after the backlash correction. }
 \label{fig:Ycorrection}
\end{figure}

The first step is to characterize the backlash magnitude along the $x$ and $y$ directions. To this purpose,  we track a fluorescent latex bead stuck on a glass plate and switch off the automatic displacement of the stage. Under these circumstances, the software recognizes the particle position but does not move the stage to keep it in the middle of the image. We start the tracking with the bead in focus in the image center and manually move the stage in $x$ and $y$ using the joystick. The bead displacement in the image should then be equal to the stage displacement. The position $(x_C, y_C)$ calculated using equation (\ref{x_chamber}) should be constant as the particle has not moved with respect to the microfluidic chamber. The apparent displacements with respect to the chamber are then solely due to the backlash error. Fig. \ref{fig:Ycorrection} (top), shows the measured $y_S$ stage position (black circles) during such a test. Using the same scale, the resulting position $y_C$ of the object in the chamber is displayed in Fig. \ref{fig:Ycorrection} (bottom). The position of the particle with respect to the microfluidic chamber, obtained from the recorded data, is not constant. It is shifted by a value of around $2\mu m$ every time the direction of the stage has been reversed and the total stage displacement has been more than this value.

To compensate for this backlash, we correct the stage position by assuming that a real displacement does not take place until it is bigger than a certain distance $dy$ (around $2\mu m$) in the case of motion inversion. This results in a delay every time the direction is reversed with a subsequent reduction of the stage displacement.  We thus correct the trajectories by shifting them to match the position where the stage was stalled. The final particle position in the channel follows from equation (\ref{x_chamber}). We repeat this procedure for different values of $dy$ and determine the value $dy^*$ that minimizes the standard deviation of the corrected position $y_C$. The correction of $y_S$ using this optimal value can be seen in Fig. \ref{fig:Ycorrection} (top, red line).  This correction also turns the step function in Fig. \ref{fig:Ycorrection} (bottom, red line) into an almost constant line that reflects the actual fixed position of the particle with respect to the chamber.

An identical procedure is performed for the $x$ direction. The optimal values $dx^*=0.45\mu m$ and $dy^*= 2.0\mu m$ are found to be identical for independent realizations of the stage displacements and have been chosen to correct the backlash in our algorithm.

\subsection{Tracking performances}

To test the quality of our tracking method we follow latex beads performing Brownian motion. \comA{Due to the intrinsic differences in the tracking principles in $x,y$ and $z$ we present two separate test. For the motion in $x$ and $y$, the main source of uncertainty comes from the presence of the backlash. Random Brownian motion is indeed a severe test for our correction due to the small persistence of the trajectories.} For the $z$ position, the main source of uncertainty is the quality of the $z$ detection algorithm.

\comA{Our first experiment is to test the backlash correction along $x$ and $y$. To do so, we follow a diffusive latex bead in a quiescent fluid by switching off the automatic displacement of the stage. Then, we follow the same particle using also the automatic stage displacement of the tracking algorithm and compare the results.}

\begin{figure}[!htb]
\begin{tabular}{l l}
  \textbf{a} & \textbf{b} \\
    \includegraphics[width=0.48\linewidth]{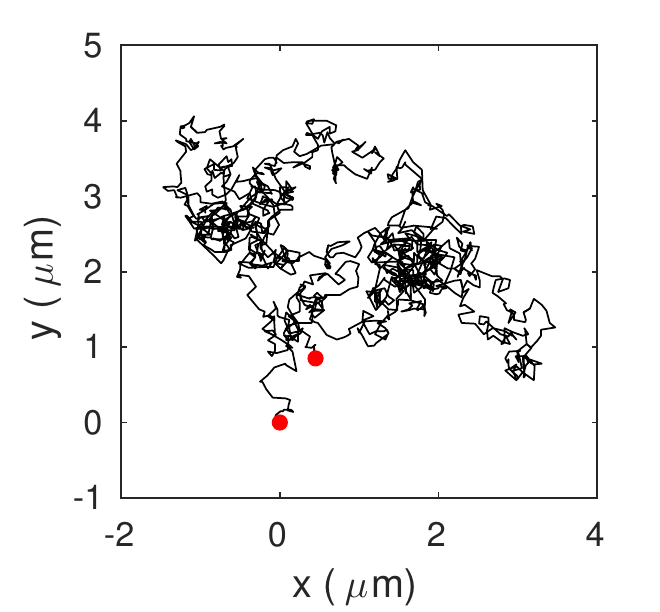} &
      \includegraphics[width=0.48\linewidth]{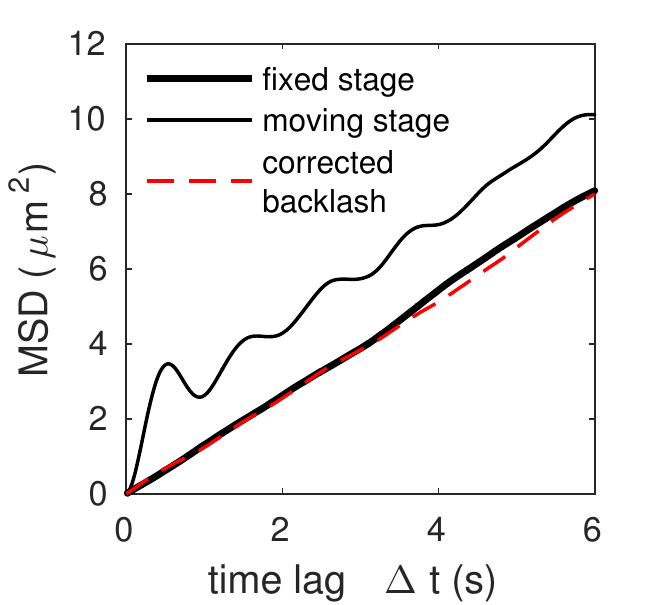}
\end{tabular}

 \caption{Tracking of a Brownian particle. (a) 2D trajectory of a latex bead undergoing Brownian motion recorded during $20s$ with an immobile stage.
 The origin of position is taken at coordinates (0,0). The beginning and the end of the trajectory are marked with red spots. (b) Mean square displacement for a trajectory of the same bead (solid thick line) with a fixed stage, with a moving stage and no backlash correction (thin solid line) and for the same trajectory after backlash correction (thick dashed line). }
 \label{fig:XYtrack}
\end{figure}

\comA{When the stage does not move,} the position of the particle is determined in real time solely by the image analysis part of the algorithm. The trajectories are then typical of a thermal Brownian diffusion with a diffusion coefficient $D_E = \frac{k_B T}{6 \pi \mu a} $ following Einstein's relation, where $a$ is the particle radius and $\mu$ the fluid viscosity. For the particle of Fig. \ref{fig:XYtrack} (a) which is a round latex particle of diameter $1.75\mu m$ from Beckman Coulter ($std = 0.02\mu m$), the theoretical diffusion coefficient in water at \comA{$26 \pm 0.5 ^oC$}  is expected to be \comA{$D_E = 0.29 \pm 0.01 \mu m^2/s$}. The mean square displacement (MSD) of the particle is defined as $\left < (\vec{r}(t+\Delta t) - \vec{r}(t))^2 \right >_t = 4 D \Delta t$ in two dimensions. The brackets denote the average along the trajectory.

To avoid the lack of statistics affecting the mean-square displacement measurements we only consider time lags smaller than one-tenth of the full track duration.  Note that the maximal time lag is relatively short for this experiment, as the tracking stops when the particles leaves the image frame. From the slope of the MSD (thick line in Fig. \ref{fig:XYtrack} (b)), we extract a diffusion coefficient $D=0.3215 \pm 0.0004 \mu m^2/s$. While this result is close to the theoretical prediction they do not agree within the error bars. This discrepancy is most likely due to the shortness of the trajectory used here. Note that the measured diffusion coefficients might differ by up to 15$\%$ for different realizations of the same experiment under these tracking conditions.

Now we track the same particle as before, but with a moving stage. The tracking procedure will keep the particle in the trapping area (see Fig. \ref{fig:setup}). In this case, the obtained trajectory is affected by the backlash. The signature of an uncorrected backlash on the tracking of a Brownian object is an offset and oscillations in the mean square displacement as seen in Fig. \ref{fig:XYtrack} (b) (thin black solid line). When the obtained trajectory is corrected with the above explained method, using the optimal values for $dx^*$ and $dy^*$, independently for each horizontal axis, the MSD curve becomes a straight line (thin red dashed line) and the diffusion coefficient measured from this corrected track is $0.3085 \pm 0.0005 \mu m^2/s$. This value differs by only 4$\%$ from the diffusion coefficient obtained from the experiment with a fixed stage and the difference is clearly within the typical scatter observed for different realizations of the same experiment. We can thus confirm that the backlash correction is satisfying.

\begin{figure}[!htb]
\begin{tabular}{l l}
  \textbf{a} & \textbf{b} \\
    \includegraphics[width=0.48\linewidth]{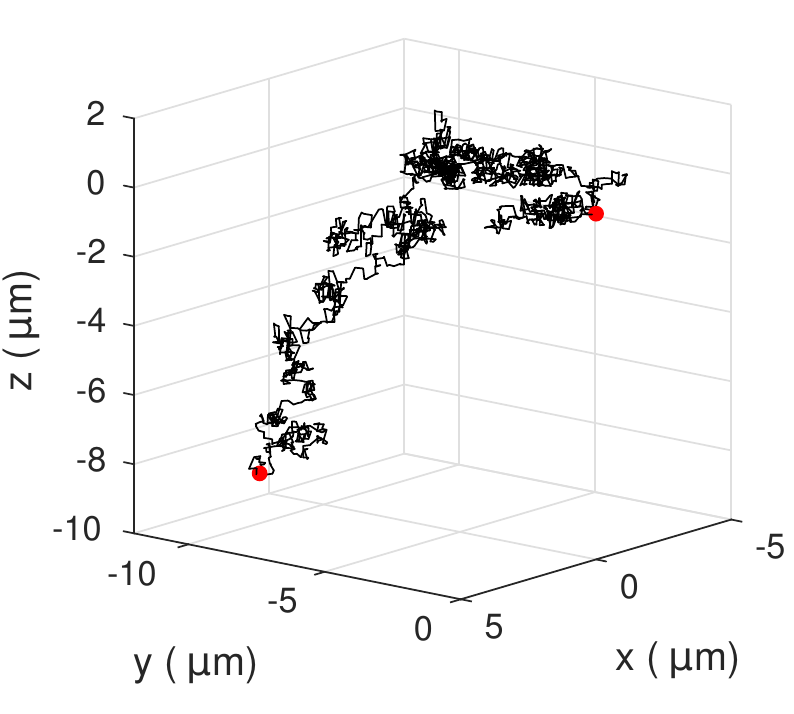} &
    \includegraphics[width=0.48\linewidth]{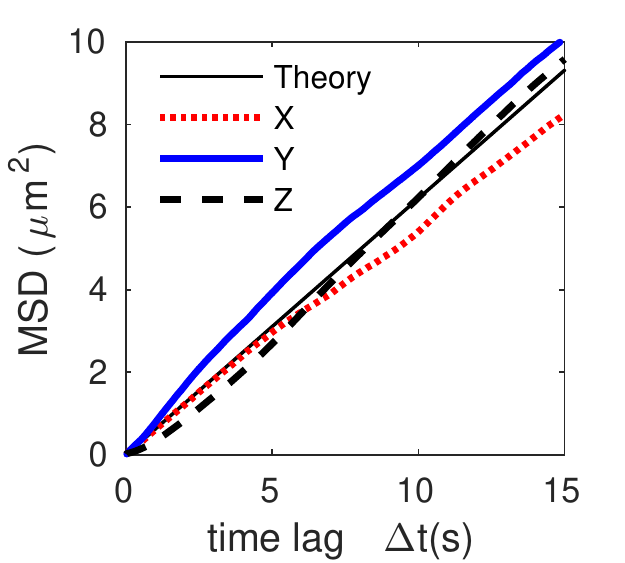}
\end{tabular}

 \caption{Tracking of a Brownian particle during sedimentation. (a) 3D trajectory during $31s$. (b) MSD by components. The $x$ and $y$ components have had backlash correction. After subtraction of the drift velocity (and backlash correction) an  agreement between the three spatial coordinates is found.}
 \label{fig:XYZtrack}
\end{figure}

\comA{Now we address the performance of the tracking in the $z$ direction.}
To this purpose, we perform a 3D tracking of a bead undergoing slow sedimentation (carboxylate particle from Polysciences of diameter $1.62 \mu m$, $\rho = 1.05g/cm^3$). The trajectory is displayed in Fig. \ref{fig:XYZtrack} (a). \comA{The theoretical sedimentation speed is $0.1 \pm 0.02 \mu m/s$. The measured sedimentation speed is $0.08 \pm 0.002 \mu m/s$, which agrees within the error bar with the theoretical value. The local velocity is subjected to thermal fluctuations and after subtraction of the average drift velocity we can calculate the MSD.}
The MSD is represented versus time lag for the three directions, after backlash correction for $x$ and $y$ in Fig. \ref{fig:XYZtrack} (b). Note that the diffusion coefficients for each of the three spatial directions separately differ slightly from each other, but agree within 10 percent. The diffusion coefficient in 3D, computed directly from one sixth of the slope of the MSD is \comA{$D = 0.312 \pm 0.0003 \mu m^2/s$} and agrees within the error bars with the theoretical value of \comA{$D_E = 0.311 \pm 0.005 \mu m^2/s$}.  The better agreement of the measured value of $D$ and the theoretical prediction $D_E$ compared to the previous experiment is most likely due to the longer duration of the track used here. From this test we can thus conclude that our z tracking method is reliable.

Note that tracking a Brownian particle is the most difficult situation for a stage in terms of the backlash, \comA{due to the small persistent time of the trajectories. The persistence time is $\approx 6.7 ms$ for the previous particle and has to be compared compared to the time resolution of our instrument which is $12.5 ms$ in $x,y$ (corresponding to 80fps). The Brownian particle will thus be in a position totally uncorrelated from the previous position at each step and subsequently the back and forth displacements of the stage perform a random walk following the particle}. The tracking of swimmers or objects under flow lead much less often to backlash, since their trajectories are more persistent. \comA{For example, typical persistence times for \textit{E. coli} bacteria are of the order of some tenth of seconds \cite{Figueroa_Thesis}}. As our method of backlash correction works well for Brownian motion, we are thus confident that it will correctly handle swimming bacteria or objects transported in a flow.

We will now address the specific uncertainties in $x,y,z$ and the time resolution. In addition to the backlash (for which we correct) the trajectories also contain additional noise introduced by the tracking procedure and the intrinsic device positioning performances. This noise can occur at various stages of the tracking process: in the detection phase when applying a local threshold to the image, during the stage motion due to the mechanical response of the stage or due to temporal delays concerning computer time and data transfer performances between the camera, the  computer and the stage.

The uncertainties in $x$ and $y$ coordinates have different values for one directional motion or for motion with repeated direction reversals, as bigger uncertainties are introduced during the backlash correction. From the noise in detection of the fixed bead used for the backlash correction we determine a standard deviation of $0.02 \mu m$ when there is no backlash at play. The backlash correction introduces an extra $x,y$ uncertainty of $0.17 \mu m$.

To determine the uncertainties in the $z$ detection, we analyze our bead during sedimentation in Fig. \ref{fig:XYZtrack}. The displacements between two consecutive images along $z$, after subtraction of the sedimentation contribution, are due to Brownian motion plus detection uncertainties in $z$. These displacements have a flat distribution centered around zero with a standard deviation of $0.3\mu m$. This distribution is wider than the distribution obtained for $x$ and $y$ from the same experiment: a normal distribution with a standard deviation of $0.09\mu m$ centered around zero. We thus conclude that the large standard deviation of $0.3\mu m$ associated with the $z$ coordinate is a measure of the uncertainty of the position determination. This uncertainty will be higher in the case of non-sperical objects randomly oriented, like microorganisms. We estimate it to be of the order of the half-mayor-semiaxes of the ellipsoid. For a typical \textit{E. coli} bacterium of body dimensions $2\mu m \times 1\mu m$ it will be $1\mu m$.

\begin{figure}
\centering    \includegraphics[width=0.9\linewidth]{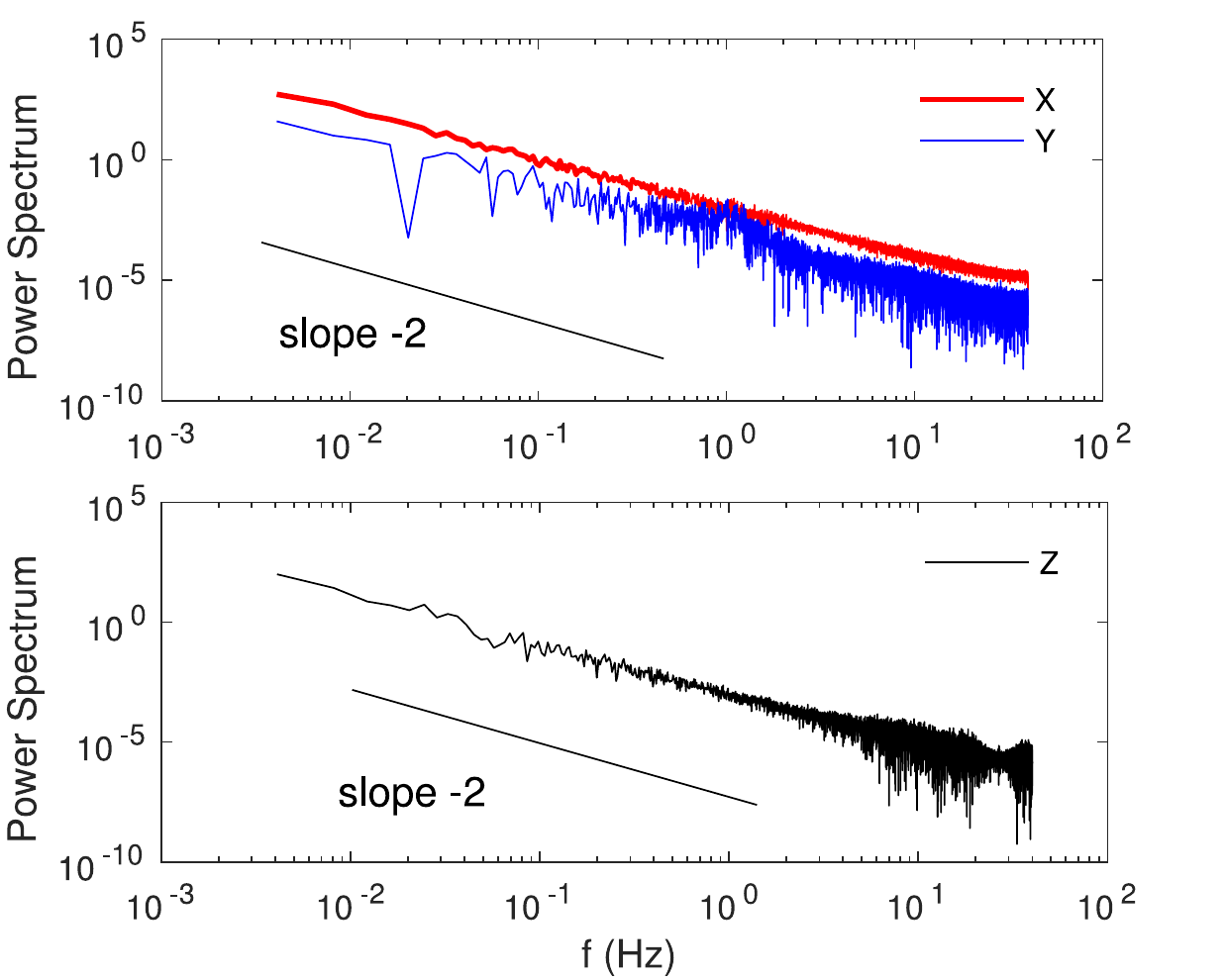}
 \caption{Power spectrum of $x$, $y$ (top) and $z$ coordinates (bottom) of the 3D trajectory of the bead of Fig. \ref{fig:XYZtrack}.}
 \label{fig:FFT_Z}
\end{figure}

To address the time resolution of the device, the spectral response of the Brownian particle of diameter $1.62 \mu m$ of Fig. \ref{fig:XYZtrack} was computed and is shown on Fig. \ref{fig:FFT_Z}. One can see the $1/f^2$ decay expected for Brownian motion. The spectrum of $y$ has a slight peak at around $1Hz$ probably introduced by the backlash correction since the raw data (before correction) do not present this feature \comA{(data not shown)}. Let us recall that the backlash amplitude in $y$ was quite large for our mechanical stage \comA{($dx^* = 0.45\mu m$, $dy^* = 2.0\mu m$)}. However, for the $x$ direction where the backlash is lesser, we do not see any trace of the correction effect. For the $z$ coordinate, the noise increases starting from a few Hertz, and has a cut-off at the effective sampling frequency $80/3 \approx 26.7 Hz$ which corresponds to the actual sampling frequency in $Z$.

The sampling frequency being $80Hz$ in $x$ and $y$ and of \comA{$26.7 Hz$} in $z$, the present instrument should allow a 3D tracking of bacteria  with a temporal resolution high enough to study the tumbling events ($0.1s \sim 8$ frames. One can can even hope to study periodic phenomena up to $40 Hz$ in $x$ and $y$, like the wobbling dynamics of a swimming bacterium \cite{liu2014helical, DiLeonardo_2017}. Moreover, we have also tested this device for tracking under flow and we were able to follow bright objects moving at the speeds up to $160\mu m/s$ in $xy$ and at around $40 \mu m/s$ along the $z$ direction.

\section{Conclusions}

In this report, we describe a 3D tracking method that can be implemented using common laboratory devices (a x,y mechanically controlled stage with a z piezo-mover, a light sensitive CCD camera, a National Instrument Trigger generator and a standard PC running a Labview Program). Such a scientific instrument has numerous applications in biology or microfluidics. With the Lagrangian tracking instrument described in this report one can follow fluorescent particles or motile micro-organisms under the microscope over large distances. The 3D tracking has a precision on the position below a micron at a rate of acquisition of several tenth of Hertz (here 80Hz). The velocity at which a particle can be tracked, for example in a flow, reaches $160\mu m/s$ horizontally and $40\mu m/s$ vertically. The performance of the method is likely to be improved with a new and faster camera of SCMOS technology flushing images on a PC running a RAID (Redundant Array of Independent Disks) and thus able to reach a higher transfer rate. The backlash problem of the mechanical stage can be corrected using an optical encoder instead of the current rotary encoder. Also the image analysis technique we propose can be adapted to track non-spherical objects or to study the microhydrodynamics of many particulate systems, like non spherical particles, flexible fibers etc., in fluids of various rheological properties.
The present set-up can also  be used to investigate the transport of motile micro-organisms like bacteria under flow in microfluidic devices and in various confined environments.

\bigskip

{\bf Acknowledgements} We thank Dr V. Martinez of Univ. Edinburgh  for useful technical discussions; Pr C. Douarche for discussions and for providing different strains of fluorescent bacteria. N. F. M. thanks support by the Pierre-Gilles de Gennes Foundation. P.B. thanks the C'Nano Ile de France for Post-doctoral financial support. We acknowledge the financial support of the ANR 2015 ``Bacflow''. A.L. and N.F.M. acknowledge support from the ERC Consolidator Grant PaDyFlow under grant agreement 682367.

%
%

\end{document}